\documentclass[aps,prl,twocolumn]{revtex4-2}
\usepackage{graphicx}
\usepackage[caption=false]{subfig}
\usepackage{amsmath}
\usepackage{amssymb}
\usepackage{bm}
\usepackage{tikz}
\usepackage{bibunits}
\def\plaq{\tikz[baseline=.1ex]{
\draw (0,0) -- (1.0ex, 0);
\draw[thin] (1.5ex,-0.5ex) node[inner sep=0pt,minimum size=0.1pt] {\scriptsize $l$};
\draw (0,0) -- (0, 1.0ex);
\draw[thin] (-0.5ex,1.5ex) node[inner sep=0pt,minimum size=0.1pt] {\scriptsize $i$};
\draw (0, 1.0ex) -- (1.0ex, 1.0ex);
\draw[thin] (1.5ex,1.5ex) node[inner sep=0pt,minimum size=0.1pt] {\scriptsize $j$};
\draw[thin] (-0.5ex,-0.5ex) node[inner sep=0pt,minimum size=0.1pt] {\scriptsize $k$};
\draw (1.0ex, 0) -- (1.0ex, 1.0ex);}
}

\usepackage{mathtools}

\DeclarePairedDelimiterX\braket[2]{\langle}{\rangle}{#1 \delimsize\vert #2}

\newcommand{\be}{\begin{equation}}
\newcommand{\ee}{\end{equation}}
\newcommand{\ba}{\begin{eqnarray}}
\newcommand{\ea}{\end{eqnarray}}

\makeatletter
\newcommand{\crossout}[1]{
  \begingroup
  \sbox\z@{#1}
  \dimen\z@=\wd\z@
  \dimen\tw@=\ht\z@
  \dimen\z@=.99626\dimen\z@   
  \dimen\tw@=.99626\dimen\tw@ 
  \edef\co@wd{\strip@pt\dimen\z@}
  \edef\co@ht{\strip@pt\dimen\tw@}
  \leavevmode
  \rlap{\pdfliteral{q 1 J 0.4 w 0 0 m \co@wd\space \co@ht\space l S Q}}
  \rlap{\pdfliteral{q 1 J 0.4 w 0 \co@ht\space m \co@wd\space 0 l S Q}}
  #1
  \endgroup
}
\makeatother

\begin{document}
\begin{bibunit}

\title{Nematic Antiferromagnetism and Deconfined Criticality from the Interplay between Electron-Phonon and Electron-Electron Interactions}
\author{ {$\rm Chao \ Wang^{1}$}, {$\rm \ Yoni \ Schattner^{1,2}$}, {$\rm \ Steven \ A. \ Kivelson^{1}$}}
\affiliation{1) Department of Physics, Stanford University, Stanford, CA 94305}
\affiliation{2) Stanford Institute for Materials and Energy Sciences, SLAC National Accelerator Laboratory and Stanford University, Menlo Park, CA 94025}

\begin{abstract}
{Systems with strong electron-phonon couplings typically exhibit various forms of charge 
order, while 
strong electron-electron interactions lead to 
magnetism.
We use determinant quantum Monte Carlo (DQMC) calculations to solve 
a model on a square lattice  
with a caricature of these interactions.
In the limit where electron-electron interactions dominate 
it has antiferromagnetic (AF) order, 
while where electron-phonon coupling dominates 
there is columnar valence-bond solid (VBS) order. 
We find 
a novel intervening phase that hosts coexisting nematic and  antiferromagnetic orders. 
We have also found evidence 
of a Landau-forbidden continuous quantum phase transition with an emergent $O(4)$ symmetry between the VBS and the nematic antiferromagnetic 
phases. 
}
\end{abstract}

\maketitle

\paragraph{\bf Introduction:}
The interplay of electron-electron and electron-phonon interactions is crucial in determining the nature of the ground state in electronic systems. While electron-electron interactions conventionally give rise to magnetism, a large electron-phonon coupling can give rise to, among other things, superconductivity, charge or bond density wave orders, as well as nematic order. 
It is also possible that the interplay between the two sorts of interactions can stabilize novel intermediate quantum phases, including spin-liquids, 
or exotic ``deconfined  quantum critical'' transitions between otherwise conventional broken-symmetry phases. 

In this work we consider 
fermions on the two dimensional square lattice 
with repulsive electron-electron interactions, as well as a coupling to local pseudo-spin degrees of freedom that are a caricature of optical phonons.  
We compute 
thermodynamic correlation functions using determinant quantum Monte Carlo (DQMC)~\cite{DQMC1, DQMC2} 
and restrict our attention to the case in which the average electron density is $n=1$ electron per site, so the simulations are free of the famous fermion minus sign problem.

\begin{figure}[!htb]
  \includegraphics[width=3.2in]{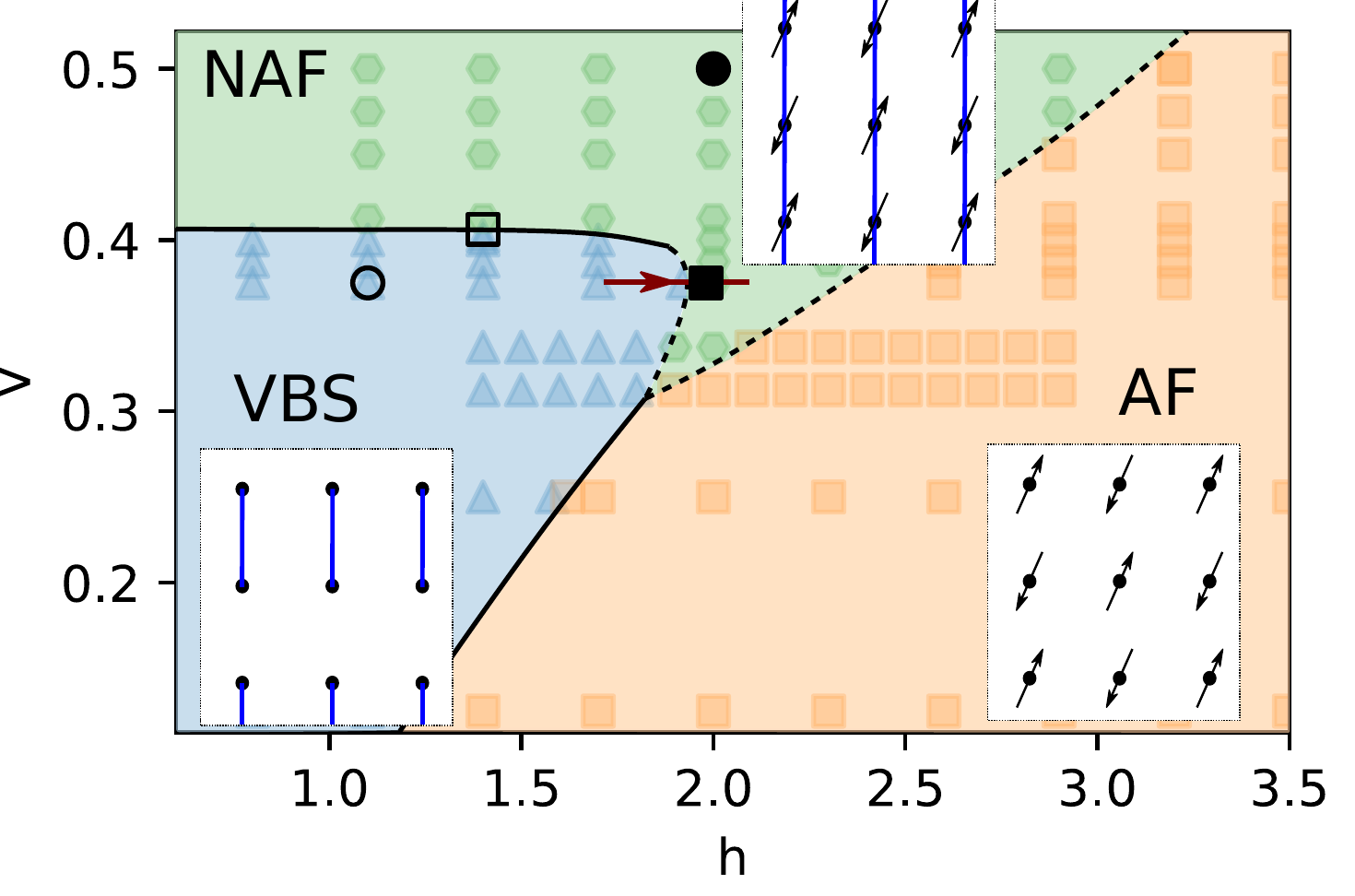}
  \caption{Phase diagram in the $h-V$ plane with $\alpha = 0.7, \tilde V=0.5$. Data points in the VBS, NAF and AF phases are shown in blue, green, and orange, respectively. The lines are a guides to the eye. Solid (dashed) lines indicate first (second) order transitions. The insets illustrate the ground state deep in the ordered phases. Blue lines represent strong bonds, and the arrows represent the spin orientations. The data presented in Figs. 2, 4(a) 4(b), and 5 correspond to the black filled circle, filled square, empty square and empty circle, respectively. Fig. 3 examines the phase transition indicated by the dark red arrow. 
  }
  \label{phasediag}
\end{figure}

Our model is conceived to exhibit two previously studied phases in extremal limits, 
an antiferromagnetic (AF) and a columnar valence-bond-solid (VBS) phase \cite{read_sachdev_89, rk} (shown in cartoons in Fig. \ref{phasediag}) both of which are incompressible at temperature $T\to 0$, and hence insulating.  
In addition to 
these two phases 
we find 
a novel insulating nematic antiferromagnetic (NAF) phase, 
in which the lattice $C_4$ rotation symmetry is broken down to $C_2$  (as it is in the VBS phase) but the only translation symmetry breaking is associated with the AF order.

There are a number of phase transitions of interest in this phase diagram. 
Most notably, 
along a portion of the phase boundary between the VBS and NAF phases (dashed line in the figure), the transition is continuous and exhibits an emergent $O(4)$ symmetry that unifies the AF and VBS order parameters. This is an example of a deconfined quantum critical point (DQCP) 
\cite{sbsvf}
at  which larger symmetries 
are predicted to emerge.  
Beyond a tricritical point, this transition becomes first order (as indicated by the solid line). 
Our results are in line with 
previous numerical studies 
\cite{loop1, loop2, loop3, sandvik_07, melko_kaul_08, okubo, dqcp_dual, dqcp_dynamical, suwa_16, two_scale, lou_09, noether_19, hong, pujari, easyplane_kagome, gazit, yuhai, ippoliti, zhao, serna, dirac_o4}
of 
various DQCPs.
In particular, 
the emergent $O(4)$ symmetry in our study is similar to that found in 
studies~\cite{dirac_o4, nematic_paramagnet} 
of the DQCP between an AF and  a Kekul\'e phase or between an AF and a nematic paramagnetic phase. 
Interestingly, in these cases the transition cannot be described in terms of the proliferation of topological point defects in either the $Z_2$ or the O(3) orders. 
In constrast, 
such a description is possible for the closely related case of a DQCP between  phases with a $U(1) $ 
and a $Z_4$ order, for which  there is good theoretical \cite{sbsvf,chong_symm_dual} and numerical evidence \cite{dqcp_dual, noether_19}
of an emergent $O(4)$ symmetry.  

More generally,
these considerations are reminiscent of earlier theories of a DQCP between VBS and AF phases, 
where duality arguments \cite{sbsvf,chong_symm_dual} suggested an $O(5)$ symmetry that unifies 
N\'eel ($O(3)$) and VBS ($Z_4$) order parameters. 
However, 
rigorous bounds on critical exponents obtained from conformal bootstrap calculations \cite{boot} 
have raised doubts about the existence of such a DQCP.
Indeed, 
elsewhere in the phase diagram we 
find a direct transition from an AF to a VBS phase, but this transition appears always to be first order - 
we find no signs of a putative DQCP with 
emergent $O(5)$ symmetry. 

Finally, there is a Landau-allowed continuous transition along the phase boundary between the AF and NAF phases.

\paragraph{\bf The Model:}
Our model is defined on the two dimensional square lattice, with electrons that interact with on-site repulsive Hubbard interactions. We also introduce pseudo-spin variables that are a caricature of phonons. Each pseudo-spin variable lives on a nearest-neighbor bond and can be thought of as representing the local lattice distortion. 

The Hamiltonian consists of three parts
\be
H=H_{\rm{e}}+ H_{\rm{ph}} + H_{\rm{int}}.
\label{model}
\ee
$H_{\rm{e}}$ is the Hubbard model for spin-$\frac{1}{2}$ fermions at half-filling
\begin{align}
    H_{\rm{e}} &= -t\sum_{\langle i,j \rangle,\sigma} \left(c_{i,\sigma}^\dagger c_{j,\sigma} + {\rm h.c.}\right) \nonumber \\
    &+ U \sum_{i} \left( n_{i,\uparrow} - \frac{1}{2}\right) \left( n_{i,\downarrow} - \frac{1}{2}\right),
\end{align}
where $i$ denotes sites and $\langle i,j \rangle$ denotes nearest-neighbors. 

$H_{\rm{ph}}$ 
is the bare phonon piece:
\begin{align}
   & H_{\rm{ph}} = - {V} \sum_{\plaq} \left(\tau^z_{ij} \tau^z_{kl} +  \tau^z_{ik} \tau^z_{jl} + \tau^z_{ij} \tau^z_{kl} \tau^z_{ik} \tau^z_{jl} \right) \nonumber \\
    &\  \ \   + ( V+\tilde V) \sum_{\langle j,i,k  \rangle} \tau^z_{ji} \tau^z_{ik}  + 6\tilde V \sum_{\langle i,j \rangle} \tau_{ij}^z - h \sum_{\langle i,j \rangle} \tau_{ij}^x,
\end{align}
where the  pseudo-spin $\vec \tau_{ij} =\vec \tau_{ji}$ represents  a two-state  ``phonon'' mode on each nearest-neighbor bond, 
the sum over $\langle j,i,k\rangle$ is over pairs of bonds with  a common vertex, the various terms proportional to $V$ and $\tilde V$  (both assumed positive) determine the favored ``classical configurations'' of the  pseudo-spins, and the transverse field $h$ gives them dynamics. 
The classical states can be visualized in a lattice-gas representation, in which a ``strong''  bond on which $\tau_{ij}^z=1$ is thought to be occupied by a dimer.  The preferred (zero energy) configurations for large $\tilde V$  
correspond to those of the hard-core dimer model~\cite{rk}, such that no two strong bonds  share a common vertex (i.e. the dimers satisfy a hard-core constraint). Positive $V$ on the other hand, favors plaquette configurations 
with exactly one pair of dimers on opposite sides. The classical ($h\to 0$)  ground-states of $H_{\rm{ph}}$ for  $\tilde V> V >0$ are the 4 symmetry-related columnar VBS states of the sort shown in the lower left inset in Fig. 1 with dimers on the blue bonds, while for  $V>\tilde V >0$, they are the 2 nematic states of the sort shown in the upper inset.

Lastly 
the electrons are coupled to 
the pseudo-spins 
by
\begin{align}
    H_{\rm{int}} = - \alpha t \sum_{\langle i,j \rangle, \sigma} \tau_{ij}^z \left[c_{i,\sigma}^\dagger c_{j,\sigma} + {\rm h.c.}\right].
\end{align}
The interactions in our model are 
such that 
the ground-state is an AF 
for $h\to \infty$ and a columnar VBS phase as $h\to 0$ so long as $V \ll \tilde V$ and  $\alpha$ is sufficiently big.  

\paragraph{\bf Calculational particulars:} In order to use the DQMC technique without encountering the sign problem, we restrict ourselves to the case of half-filling of fermions. We apply a discrete Hubbard-Stratonovich decoupling in the spin channel to represent the Hubbard interaction \cite{supplemental}. We have performed DQMC simulations at finite temperatures, with imaginary time discretization $\Delta \tau=0.1$ and systems of linear size up to $L=18$ and down to temperatures $T=1/18$. Throughout this letter we use periodic boundary conditions, and fix $t=1, U=3, \tilde V=0.5$, and $\alpha=0.7$ unless mentioned otherwise.  As illustrated in Fig. 1, we then explore the zero temperature phase diagram (by extrapolating finite $T$ results to $T=0$) as a function of $h$ and $V$.

Where $C_4$ symmetry is spontaneously broken, in order to avoid complications due to metastable domain structures, we typically seed our DQMC runs with a configuration obtained by introducing an explicit symmetry breaking field for an initial 3000 DQMC steps, but then removing this field so that the model has the requisite $C_4$ symmetry for all subsequent steps.
We illustrate our most significant findings with representative figures in the main text, but present more complete data in the Supplemental Material.

\begin{figure}[htbp!]
    \captionsetup[subfigure]{justification=centering}
    \centering
    \includegraphics[width=3.4in]{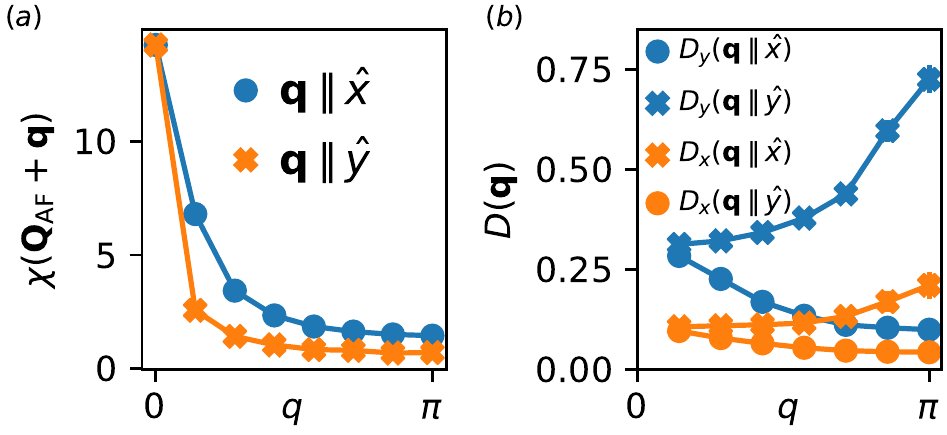}
    \caption{Correlation functions in momentum space $\mathbf{q}$ in the NAF phase ($h=2.0, V=0.5$ - the black filled circle in Fig. 1 - with $L=\beta=14$). 
    (a) Static spin-spin susceptibility as a function of deviation from AF ordering vector $\mathbf{Q}_{\rm AF} = (\pi,\pi)$. (b) Static bond-bond susceptibility for $x$- and $y$-direction bonds.}
    \label{fig:naf}
\end{figure}

\paragraph{\bf Nematic antiferromagnet:} 
Fig. \ref{phasediag} shows that the NAF 
arises as an 
intermediate phase. 
In Fig. \ref{fig:naf} we 
show the spin 
and pseudo-spin
susceptibilities as functions of momentum ($\mathbf{q}$) at a representative point in this phase. The susceptibilities are defined as
\begin{align}
    \chi(\mathbf{q}) &= \frac{1}{L^2} \sum_{i, j} \int_{0}^{\beta} d\tau \ \langle \vec{S}_i(\tau) \cdot \vec{S}_j(0) \rangle e^{i \mathbf{q} \cdot (\mathbf{r}_i - \mathbf{r}_j)} \label{eqn:spin-spin} \\
    D_a(\mathbf{q}) &= \frac{1}{L^2} \sum_{\langle i,j \rangle_{a} \langle k,l \rangle_{a}} \int_{0}^{\beta} d\tau \ \langle \tau_{ij}^z(\tau) \tau_{kl}^z(0) \rangle e^{i \mathbf{q} \cdot (\mathbf{r}_{ij} - \mathbf{r}_{kl})}, \label{eqn:bond-bond}
\end{align}
where $\vec{S}_i$ is the electron spin, $\mathbf{r}_{ij}$ is the position of the center of the bond 
$\langle i,j\rangle$, and the subscript $a= x$ or $y$ signifies the orientation of the bond.
That this is a magnetically ordered state is shown by the presence of a Bragg peak in $\chi(\mathbf{q})$ as $T\to 0$, as is evident in the figure and which we have corroborated by finite size scaling.  The fact that both correlation functions depend on the direction of $\mathbf{q}$ shows that the state spontaneously breaks $C_4$ symmetry down to $C_2$, i.e. that it is nematic.

\paragraph{\bf VBS to NAF transition:}
As shown in Fig. \ref{phasediag}, 
as a function of increasing $V$ at fixed small $h$, 
there is a strongly first order transition (indicated by a solid line) from a VBS to the NAF, characterized by a discontinuous jump in first derivatives of free energy \cite{supplemental}. 
Indeed, since  the NAF spontaneously breaks time-reversal (TR) symmetry but  is invariant under the product of translation  and time-reversal (TrTR) while the VBS preserves TR but breaks TrTR, conventional Landau theory implies the transition must be first order.
However, at larger $h$, the phase boundary bends sharply and,
beyond a tri-critical point, the transition becomes continuous (dashed line) within our numerical resolution. 
Eventually 
the phase boundary ends at a bi-critical point. 
(We have not yet explored these multicritical points in detail, but we note that since the phase boundary that links them is exotic, they may have unusual features as well.)

We now focus on the continuous 
VBS-NAF transitions. 
To be concrete, we fix $V=0.375$ and study the finite size scaling behavior of the static AF and VBS susceptibilities $\chi_{\rm AF}, \chi_{\rm VBS}$ as a function of $h$. 
These are 
given by the expressions in Eqn. \ref{eqn:spin-spin} and \ref{eqn:bond-bond} evaluated at $\mathbf{Q}_\mathrm{AF}=(\pi,\pi)$ and $\mathbf{Q}_\mathrm{VBS}=(0,\pi)$, respectively, with the bond direction set to $a=y$.
Since both the VBS and the NAF phases break the $C_4$ rotational symmetry, in the 
remaining calcuations reported here we have applied a small explicit $C_4$ symmetry breaking in our simulations to stabilize our results, by making the hopping matrix element $t$ to be slightly different in the two directions (with $t_x=0.97, t_y=1.0$) \cite{shift_critical_point}.
As a consequence, the pseudo-spins correlations are stronger for $y$-directed bonds, which we will refer to as the nematic direction. The nematic direction also corresponds to the direction of the ordering wavevector in the VBS state. (See inset to Fig. \ref{phasediag}.) Note that in the presence of this explicit symmetry breaking, the VBS order
has $Z_2$ (Ising) character, corresponding to the breaking of translational symmetry.

Assuming the transition is continuous, on theoretical grounds we expect conformal symmetry with dynamical critical exponent $z=1$.  
We thus scale space and time together by taking $\beta=L$ 
and express the susceptibilities 
in the
scaling forms (neglecting corrections to scaling)
\begin{align}
    \chi(L) &= L^{d+1-\eta}\ \tilde{\chi}\left[(h-h_c) L^{1/\nu}\right],
\end{align}
where 
$d=2$, 
$\chi$ 
is either $\chi_{\rm{AF}}$ or $\chi_{\rm{VBS}}$,
and $\nu$ is the 
 correlation length 
 and $\eta$ 
 the anomalous exponent.
\begin{figure}[!htb]  
  \includegraphics[width=3.4in]{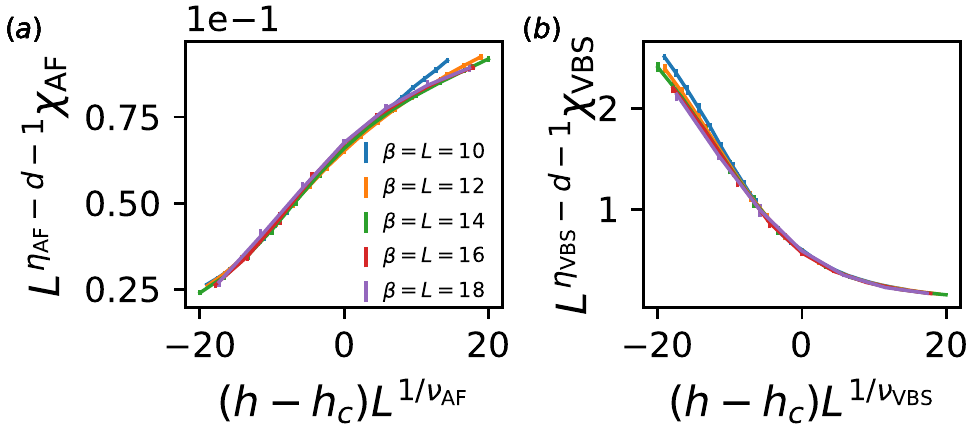}
  \caption{Finite size scaling collapse of AF and VBS susceptibilities along the dark red arrow in Fig. 1 ($V=0.375$). The location of the transition is identified as $h_c \approx 1.98$. The critical exponents for the two different orders are within error bars from each other.}
  \label{scaling-vbs-naf}
\end{figure}

In Fig.~\ref{scaling-vbs-naf} we show the finite-size scaling collapse results for these two susceptibilities taking $h_c \approx 1.98$. We obtain 
 $1/\nu_{\rm AF} = 2.2 \pm 0.4, 1/\nu_{\rm VBS} = 2.0 \pm 0.4$ and $\eta_{\rm AF} = 0.65 \pm 0.2, \eta_{\rm VBS} = 0.65 \pm 0.2$. These results indicate a direct and continuous transition between the two phases. The 
 near equivalence of the exponents extracted from the AF and VBS susceptibilities hints at an emergent $O(4)$ symmetry that unifies the three components of the spin AF with the single component VBS. The Binder ratios close to the transition provides further support for the continuous nature of the transition \cite{supplemental}.

\begin{figure}[htbp!]  
    \captionsetup[subfigure]{justification=centering}
    \centering
    \subfloat{
        \centering
        \includegraphics[width=0.23\textwidth]{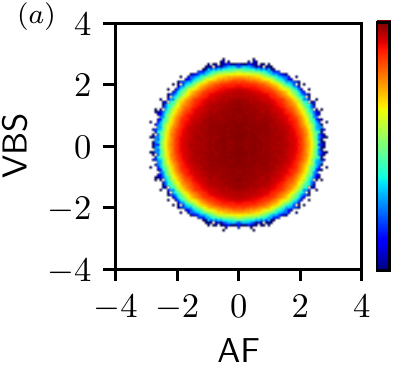}
    }
    \subfloat{
        \centering
        \includegraphics[width=0.23\textwidth]{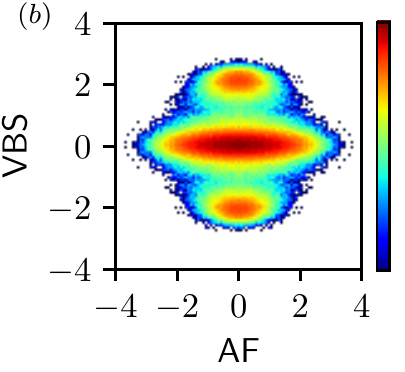}
    }
    \caption{Histograms for the joint distribution of normalized AF and VBS order parameters $({\phi}_{\rm AF}^3, {\phi}_{\rm VBS}^y)$. (a): At a continuous transition between VBS and NAF ($h=1.98, V = 0.375$ - the filled square in Fig. 1). (b): At a first order transition between VBS and NAF ($h=1.4, V = 0.40625$ - the empty square in Fig. 1).  In both cases, $L=\beta=16$.}
    \label{histogram}
\end{figure}

\paragraph{\bf Emergent $O(4)$ Symmetry:}
To further investigate the possibility of a larger emergent symmetry at the critical point, we examine the relation between 
one component of the AF order parameter $\vec{\phi}_{\rm AF}$ 
and the $y$-component of VBS order parameter $\phi_{\rm VBS}$, defined as:
\begin{align}
    &\vec{\phi}_{\rm AF} \equiv \frac{1}{{\cal N}_{\mathrm{AF}}} \int_0^\beta d\tau \sum_i \vec{S}_i(\tau) e^{i \mathbf{Q}_{\rm AF} \cdot \mathbf{r}_i},\\
    &\phi_{\rm VBS}^y \equiv \frac{1}{{\cal N}_{\rm{VBS}}} \int_0^\beta d\tau \sum_{\langle i, j \rangle_y} \tau^z_{ij}(\tau) e^{i \mathbf{Q}_{\rm VBS} \cdot \mathbf{r}_{ij}},
\end{align}
where the normalization factors ${\cal N}_{\rm AF}=\sqrt{\beta L^2 \chi_\mathrm{AF} / 3}$, ${\cal N}_{\rm VBS}=\sqrt{\beta L^2 \chi_\mathrm{VBS}}$ are defined so that $\langle \left[\phi_{\rm AF}^a \right]^2\rangle=1$ for $a=1,2,3$ and $\langle \left[\phi_{\rm VBS}^y \right]^2\rangle=1$.
In Fig. \ref{histogram} (a) we present a histogram of the joint probability distribution of $({\phi}_{\rm AF}^3, {\phi}_{\rm VBS}^y)$ at the critical point, $h$, 
where the obvious rotational symmetry serves to visualize this emergent symmetry.  For comparison, in Fig. \ref{histogram} (b) we show the analogous histogram at the point of a first order the transition between the same two phase, where only the explicit $Z_2 \times Z_2$ symmetry is present.

It is 
an established check to examine the $O(4)$ non-invariant moments \cite{loop2}  
$F_n \equiv \langle\phi^4 \cos(n\theta)\rangle$ for $n=2$ and 4 - in polar coordinates,  $\phi e^{i\theta} = \phi_{\rm AF}^3+i\phi_{\rm VBS}^y$. 
Vanishing moments imply an $O(4)$ symmetry.
In Table I we observe that at $h=1.98$ near the critical point $h_c$, the values for the two moments extrapolate to zero in the thermodynamic and zero temperature limit, strongly indicative of emergent $O(4)$ symmetry. 

\begin{table}[h]
\centering
\begin{tabular}{|l|l|l|}
\hline 
  & $F_2$  & $F_4$ \\ \hline
 $L=\beta=10$ & $0.28 \pm 0.04$  &  $-0.09 \pm 0.03$ \\ \hline
 $L=\beta=12$ & $0.20 \pm 0.05$  & $-0.08 \pm 0.03$  \\ \hline
 $L=\beta=14$ & $0.16 \pm 0.05$  & $0.06 \pm 0.05$  \\ \hline
 $L=\beta=16$ & $0.09 \pm 0.02$  & $0.05 \pm 0.03$ \\ \hline
\end{tabular}
\caption{Values for $O(4)$ non-invariant moments at $h=1.98$ - solid square in Fig. 1. Both moments are consistent with zero when extrapolated to the thermodynamic and zero temperature limits.}
\label{tbl:fs}
\end{table}

\paragraph{\bf Topological defects:}
From 
the effective field theory perspective, such an 
unconventional (``deconfined'') quantum critical point can be described in terms of a non-linear Sigma model (NLSM) 
with a four-component order parameter (1 for the VBS and 3 for AF orders) augmented by a 2+1 dimensional
$\theta$-topological term~\cite{sbsvf,chong_symm_dual}. 
The  $\theta$-
term 
connects the AF and VBS orders, so that 
even away from criticality, one expects the subdominant order parameter to appear 
where the dominant order is suppressed - especially 
near 
topological defects.
Since point-like topological defects 
do not arise for the relevant symmetries,  we focus on line defects.

\begin{figure}[!htb]  
  \includegraphics[width=2.6in]{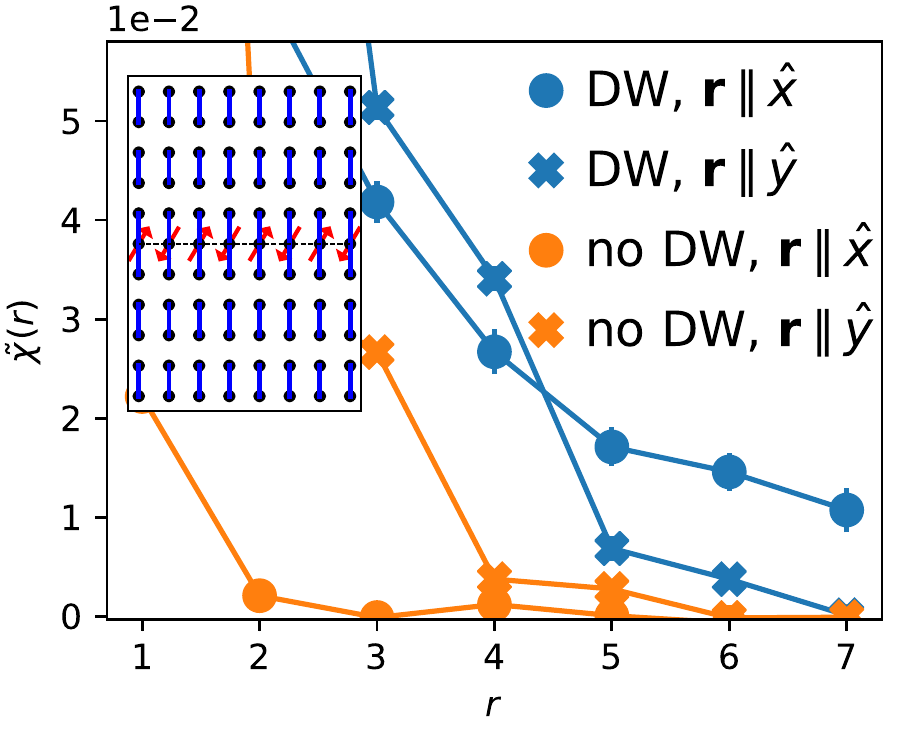}
  \caption{Real space AF correlations $\tilde \chi(\mathbf r)$ in the VBS phase with ($L_y = 15$) and without ($L_y=14$) a domain wall (DW) in the VBS order parameter. Here, $h=1.1, V=0.375$, $\beta=16$ and $L_x=14$, indicated by an empty circle in Fig. 1.
  The inset is a cartoon of the pseudo-spin configuration with a domain-wall (marked by dashed line), along which AF order is strongly enhanced.} 
  \label{domain-wall-af}
\end{figure}

At a domain-wall of the VBS order, where ${\phi}_{\rm VBS}^{y}$ changes sign (and thus passes through zero) 
we thus expect that quasi-long range AF order should develop 
along the VBS domain-wall. In our DQMC study, we can introduce a domain-wall by having an odd number of sites along one of the spatial directions. 
We examine the real-space version of the AF susceptibility (from Eq. 5)
\begin{align}
    \tilde \chi(\mathbf{r}) = \frac{1}{L_x L_y} \sum_{i} \int_{0}^{\beta} d\tau \langle \vec{S}(\mathbf{r}_i, \tau) \cdot \vec{S}(\mathbf{r}_i+\mathbf{r}, 0) \rangle e^{i \mathbf{Q}_{\rm AF} \cdot \mathbf{r}},
\end{align}
where $L_x$ sites is the number of sites in the $x$-direction and $L_y$ the number of sites in the $y$-direction. We will consider even values of $L_x$, and $\mathbf{Q}_{\rm AF}=(\pi,\pi+\delta)$, where $\delta$=0 when $L_x$ is even, and $\delta=-\pi/L_y$ when $L_y$ is odd. When $L_y$ is odd
, the VBS order is forced to have a domain wall along the $x$-direction. We perform such an experiment at $h=1.1, V=0.375$, 
well in the VBS phase. As shown in Fig. \ref{domain-wall-af}, 
when $L_x = 14, L_y = 15$, the long-distance AF correlation is strongly enhanced in the $x$-direction, along which the domain wall is aligned, 
while in the $L_x = L_y = 14$ case, the AF correlations are short-ranged \cite{dw_supplemental}.

\paragraph{\bf Discussions and Conclusions:}

It is worth mentioning several things we had hoped to see but failed to find in the present study.  Nowhere in the range of parameters we have explored did we find an intermediate spin-liquid phase between the AF and VBS phases, despite superficial similarities between our model and the quantum dimer model.  Moreover, for some parameter ranges, there is a direct transition between the AF and VBS phases, but it appears always to be first order, although sometimes weakly so - nowhere did we observe a deconfined quantum critical transition with emergent O(5) symmetry.  
We also failed to observe a superconducting phase, which one might suspect could emerge at large $h$ due to the effective attraction induced by the phonons.
Note also in Fig. \ref{phasediag} that the continuous line of deconfined quantum transition between the VBS and NAF ends at two multi-critical points. 
These may have exotic properties that we hope to explore in future work.

Some time ago,  
 a nearly continuous transition between VBS and AF phases 
was observed (using NMR \cite{stuart}) in $\rm (TMTTF)_2PF_6$, a 
quasi-2D system with orthorhombic symmetry. Since the $C_4$ symmetry is already explicitly broken, this transition 
is equivalent to 
a VBS to NAF transition.  It was speculated that its nearly continuous nature may be related to a lower dimensional (two-dimensional) quantum critical point, 
which, if true, would be expected on the basis of the present analysis to exhibit an emergent O(4) symmetry.

\begin{acknowledgments}  We acknowledge useful discussions with  H. Yao and S. Brown.
This work was supported in part by the Department of Energy, Office of Science, Basic Energy Sciences, Material Sciences and Engineering Division, under contract No. DE-AC02-76SF00515 (SAK), and also in part by the National Science Foundation (NSF) under Grant No. DMR2000987 (CW and YS). YS was also supported by the Gordon and Betty Moore Foundation’s EPiQS Initiative through Grant GBMF4302 and GBMF8686, and by the Zuckerman STEM Leadership Program. The numerical simulations were performed on the Sherlock cluster at Stanford University.
\end{acknowledgments}

\begin{appendix}

\end{appendix}
\end{bibunit}

\begin{thebibliography}{100}
\bibitem{shift_critical_point} The small explicit symmetry breaking field shifts the position of the critical points slightly, but does not change their properties.
\end{thebibliography}
\end{document}


\begin{bibunit}

\title{Supplemental Material: Nematic Antiferromagnetism and Deconfined Criticality from the Interplay between Electron-Phonon and Electron-Electron Interactions}
\author{ {$\rm Chao \ Wang^{1}$}, {$\rm \ Yoni \ Schattner^{1,2}$}, {$\rm \ Steven \ A. \ Kivelson^{1}$}}
\affiliation{1) Department of Physics, Stanford University, Stanford, CA 94305}
\affiliation{2) Stanford Institute for Materials and Energy Sciences, SLAC National Accelerator Laboratory and Stanford University, Menlo Park, CA 94025}

\maketitle

\setcounter{figure}{0}
\onecolumngrid
\setcounter{equation}{0}

\section{Technical aspects of the DQMC simulations}
To prepare the model for a DQMC simulation we write down the partition function in discrete imaginary time,
\begin{equation}
Z=\mathrm{Tr} e^{-\beta H} = \mathrm{Tr} \prod_{n=1}^{N/2} (\hat{B} \hat{B}^{\dagger}) + O(\Delta \tau^2),
\end{equation}
where $\beta=N\Delta \tau$ and
\begin{equation} 
\hat B = e^{-\Delta\tau H_U} e^{-\Delta\tau H_h /2} e^{-\Delta\tau H_{V,\tilde V}} \left(\prod_{m=1}^4 e^{-\Delta\tau \hat K_m}\right) e^{-\Delta\tau H_h /2}.
\end{equation}
Here, $H_U= U \frac{\partial H}{\partial U}$ is the Hubbard interaction term,
$H_h= h \frac{\partial H}{\partial h}$ is the transverse field term,
$H_{V,\tilde V} = V \frac{\partial H}{\partial V}+\tilde V \frac{\partial H}{\partial \tilde V}$ contains the $V$ and $\tilde V$ terms. 
The kinetic energy part of the Hamiltonian is decomposed into four groups such that $\langle i,j\rangle_m$ with $m=1,2,3,4$ enumerates the horizontal or vertical bonds originating from a site with an even or odd index along the bond direction, so that 
\begin{equation}
\hat K_m = \sum_{\sigma, \langle i,j\rangle_m} -t_{ij} (1+\alpha \tau_{ij}^z) c_{i\sigma}^\dagger c_{j\sigma} + \mathrm{H.c}.    
\end{equation}

We decouple the Hubbard term using the discrete Hubbard-Stratonovich transformation in the spin channel \cite{Hirsch}. Namely,
\begin{align}
    \exp (-\Delta \tau H_U) &= C \sum_{\eta_i = \pm 1}  \exp \big(\lambda \sum_i \eta_i \left( n_{i, \uparrow} - n_{i, \downarrow} \right) \big),
    \label{hs1}
\end{align}
where $n_{i,\sigma} = c_{i, \sigma}^{\dagger} c_{i, \sigma}$, $\eta_i$ are Hubbard-Stratonovich (HS) fields living on sites, and $\cosh(\lambda) = \exp(\Delta \tau U/2)$, and $C$ is an overall constant.

Next, we insert the identity operator in the $\tau^z$ basis between every $e^{-\Delta \tau \hat H_{V,\tilde V}}$ operator in the partition function, and trace over the fermions, to obtain
\begin{equation}
Z\propto \sum_{\tau_{ij;n}=\pm 1} \sum_{\eta_{i;n}=\pm 1}e^{-S_{\mathrm{ph}}} \det \left(G^{-1}\right),
\end{equation}
where $G$ is the fermionic Green's function, which depends implicitly on the configuration of $\eta$ and $\tau$ \cite{Assaad, nematicQMC}.
The phonon part of the action is given by
\begin{align}
    S_{\rm ph} &= \frac{1}{2} \log [\tanh(\Delta \tau h)] \sum_{\langle i,j \rangle, n} \tau_{ij;n} \tau_{ij;n+1} -{V} \sum_{\plaq, n} \left(\tau_{ij;n} \tau_{kl;n} +  \tau_{ik;n} \tau_{jl;n} + \tau_{ij;n} \tau_{kl;n} \tau_{ik;n} \tau_{jl;n} \right) \nonumber \\
    &+ (V+\tilde V) \sum_{\langle j,i,k  \rangle, n} \tau_{ji;n} \tau_{ik;n} + 6\tilde V \sum_{\langle i,j \rangle, n} \tau_{ij;n}.
\end{align}

At half filling and with nearest neighbour hoppings, for every configuration of the phonon and Hubbard-stratonovich fields, the model is symmetric under the the combination of a particle-hole and time reversal operation  $\mathcal{A} = \mathcal{C T}$.
Here $\mathcal C$ is a particle hole transformation, under which $c_{i\sigma}\rightarrow (-1)^{x_i + y_i} c^\dagger_{i\sigma}$, and $\mathcal {T}$ is the usual time reversal operator $\mathcal{T}= i S_y K$. The antiunitary $\mathcal {A}$ satisfies $\mathcal{A}^2=-1$, thereby guaranteeing the absence of the sign problem \cite{sufficient}.

Note that by the same argument, the model remains sign-free even in the presence of an orbital magnetic field. Applying a weak magnetic field by threading a single flux quantum through the system is known to alleviate spurious finite size effects at low temperatures \cite{field_finite_size}, and that is what we have done throughout the paper. 

To simplify calculations, we extract the spin-spin correlations from the correlations of the Hubbard-Stratonovich fields $\eta$. It can be shown that  
\begin{equation}
\chi_{\mathrm{AF}}(\mathbf{q})=
3\Delta\tau \chi_{\eta}(\mathbf{q}) + \Delta \chi,
\label{chiAF}
\end{equation}
where
\begin{equation}
    \chi_\eta(\mathbf q) = \frac{F}{NL^2} \sum_{ij} \sum_{nn'} \langle \eta_{i,n} \eta_{j,n'}\rangle e^{i\mathbf{q} \cdot (\mathbf r_i -\mathbf r_j)},
\end{equation}
$F=\frac{1}{4}\frac{1}{1-e^{-\Delta\tau U}}$ and $\Delta\chi = \frac{\Delta\tau}{L^2}\sum_i \left(\langle {\vec S_i}^2\rangle - 3 F\right)$. This relation can be understood by applying the mapping
\begin{equation}
    S^z_i(\tau=n\Delta\tau) \rightarrow \sqrt F \eta_{i,n},
    \label{eta_mapping}
\end{equation} which is exact for $N$ point correlation functions, so long as no two operators share the same site and time index. 
In Eqn. \ref{chiAF} such ``contact terms" are corrected for in $\Delta\chi$, and are small by factors of $L^{-\gamma/\nu}$ close to the AF critical point, where $\gamma$ and $\nu$ are the conventional critical exponents.
A similar relation can be found for the fourth order correlator, however, since we are only focused on the vicinity of AF transitions, we neglect the analogous contact terms, and for the sake of simplicity use the mapping Eqn. \ref{eta_mapping}. We have checked that the systematic error induced by this procedure is smaller than the statistical error that is already present in our simulations.

We conclude this technical summary by mentioning that in order to efficiently utilize all our DQMC simulation data,
we employ a standard reweighting method \cite{reweight} to combine data taken at slightly different values of $h$, holding all other parameters fixed.

\section{Valence-bond solid to nematic antiferromagnet transition}
Here we provide more details regarding the continuous transition between the valence-bond solid (VBS) and nematic antiferromagnetic (NAF) phases, described in the main text. We consider the antiferromagnetic (AF) and VBS (along $y$-direction) order parameters $\vec \phi_{\mathrm AF}$, $\phi_{\rm VBS} ^y$ as defined in Eqn. 8 and 9 in the main text.
We examine their Binder ratios \cite{binder0, binder0.1}
$R_{4, \rm AF} = \langle[\phi_{\rm{AF}}^3]^4\rangle$ and $R_{4, \rm{VBS}} = \langle[\phi_{\rm VBS}^y]^4\rangle$.

The Binder ratios play an important role in the theory of finite size scaling close to phase transitions. At a continuous transition, as the system size $L$ approaches the thermodynamic limit, the curves of $R_4$ vs. the tuning parameter cross at the critical point.

The Binder ratio is also important for the identification of first-order transitions. When the correlation lengths are finite but comparable with computationally accessible system sizes, distinguishing between weakly first-order transitions and continuous transitions becomes a difficult task. In such situations, finite size scaling collapse, consistent with continuous transitions, may be misleading. To this end, the Binder ratio serves as a useful tool. It is well-known \cite{binder1, binder2, sandvik_first_order} that at a first-order transition the binder ratio exhibits a divergent behavior near the transition as system size $L$ tends towards infinity. More specifically, the Binder ratio has a peak that occurs at $|h-h_c| \sim L^{-d} \ln L$, and the peak should be volume-divergent, i.e. $\sim L^d$ ($d$ is spatial dimension). It was demonstrated \cite{sandvik_first_order} in the context of $q$-state Potts models that for weakly first-order transitions, a peak in the Binder ratio that grows with system size $L$ can emerge even when finite-size scaling of susceptibilities for the numerically accessible range of $L$ suggests continuous transitions.

\begin{figure}[!htb] 
  \includegraphics[width=5.0in]{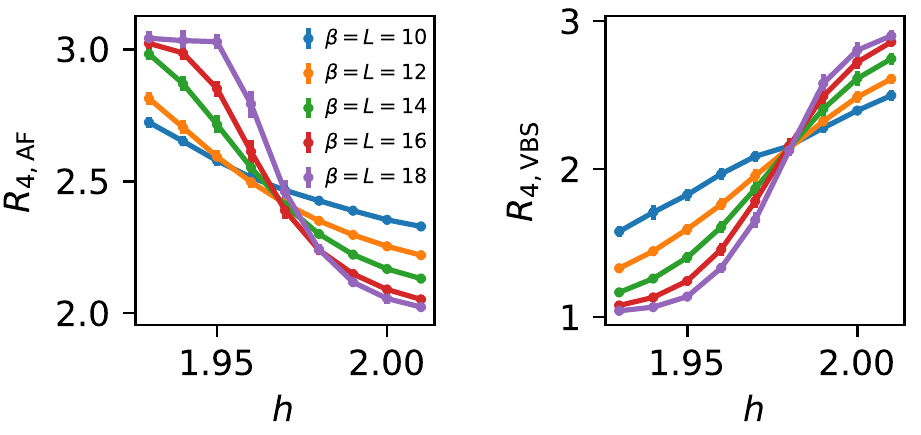}
  \caption{The Binder ratios for AF and VBS order parameters. The crossing point for AF converges to $h \approx 1.98$ as $\beta=L$ increases. The crossing point for VBS occurs at $h \approx 1.98$.}
  \label{binder-vbs-naf}
\end{figure}

\paragraph{Continuous transition}
To provide further evidence for the existence of a continous VBS-NAF transition, we plot the Binder ratios for the two order parameters in Fig. \ref{binder-vbs-naf} for various values of $L$. Here we  scale $\beta=L$, fix $V=0.375$ and use slightly anisotropic hopping parameters $t_x=0.97, t_y=1$.
First, note the absence of any signatures of a diverging peak in the Binder cumulant near the transition. By considering $h_{e}(L)$, the crossing point between $R_4(L)$ and $R_4(L+2)$, we observe that for both order parameters, $h_{e}(L)$ approaches $1.98$ at the largest system sizes available to us.
This supports the proposition that there is a single continuous transition at $h_c \approx 1.98$.

\begin{figure}[!htb] 
  \includegraphics[width=5.0in]{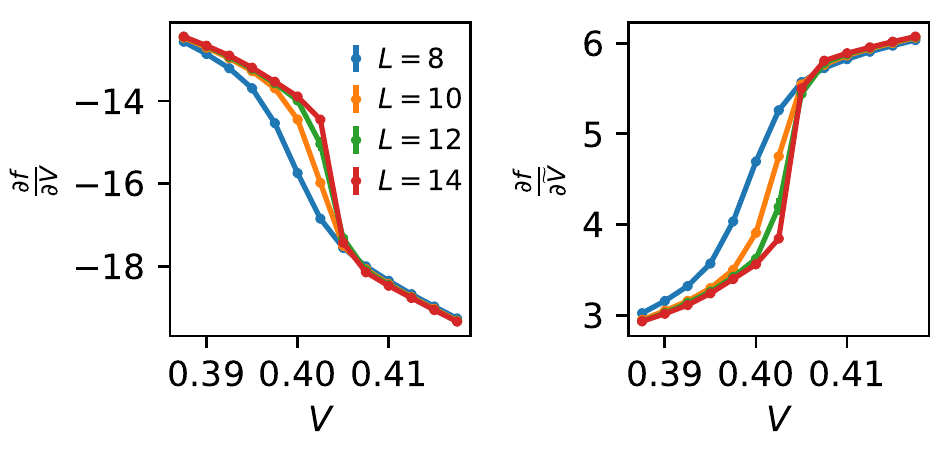}
  \caption{First derivatives of free energy density for various system sizes, across a transition from VBS to nematic, with $\beta=12.0$. They exhibit a discontinuous jump at the transition near $V_c \approx 0.4025$, which indicate that the transition is first order.}
  \label{derivative-vbs-naf-discontinuous}
\end{figure}

\paragraph{First-order transition}
We now take a different cut through the phase diagram, and show that the finite temperature transition to the VBS state is first order. Fig. \ref{derivative-vbs-naf-discontinuous} shows two first derivatives of the free energy density, which feature significant discontinuous jumps, indicating that the transition is discontinuous. Here we use isotropic hopping parameters $t_x=t_y=1$, set $h=1.4$, $\beta=12$ and vary $V$. 

\section{Domain walls in the valence bond solid phase}
\begin{figure}[!htb] 
  \includegraphics[width=2.5in]{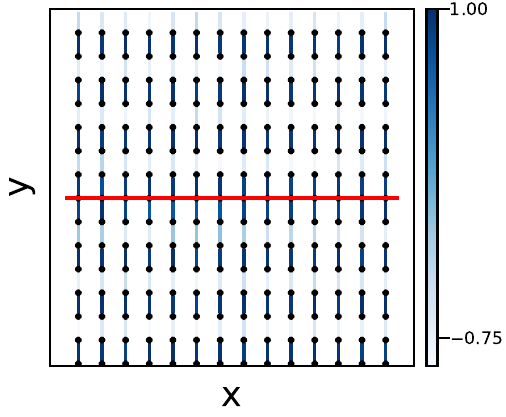}
  \caption{A typical configuration of pseudo-spins in the presence of a domain wall in the VBS phase.
  The value of $\tau^z$ for y directed bonds, averaged over imaginary time, is encoded in the color of the blue lines. The domain wall (marked by a red line), lies along the $x$-direction. 
  The parameters used here correspond to Fig. 5 in the main text, namely $L_x=14,L_y=15$, $h=1.1, V=0.375, \beta=16$.}
  \label{domain-wall-suppl}
\end{figure}

As discussed in the main text, we introduce a domain wall in the VBS order, perpendicular to the nematic direction ($y$-direction), by setting the number of sites in the $y$ direction to be odd. In Fig. \ref{domain-wall-suppl} we present direct evidence for the existence of the domain wall in a typical configuration of the pseudo-spins.

\section{Nematic-antiferromagnet to antiferromagnet transition}
\begin{figure}[!htb] 
  \includegraphics[width=2.5in]{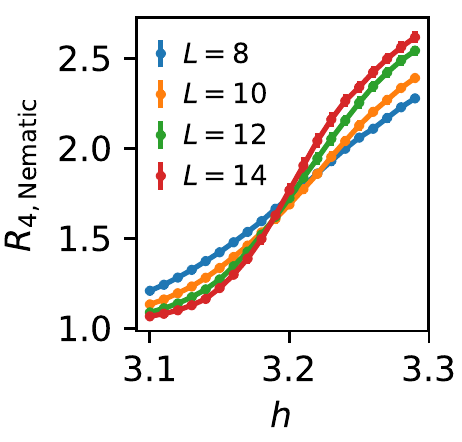}
  \caption{The Binder ratios for the nematic order parameter across the NAF to AF transition, at $\beta=10.0$. The crossing point for AF converges to $h \approx 3.19$ as $L$ increases. There is no signature of a divergent peak near the transition.}
  \label{binder-naf-to-af}
\end{figure}

As mentioned in the main text, the transition between NAF and AF phases can be continuous within conventional Landau theory. To illustrate that this is indeed the case in our model, we study the transition into the nematic state at a finite, low temperature $T=0.1$. Fig. \ref{binder-naf-to-af} shows the Binder ratio for the nematic order parameter $R_{4, {\rm Nematic}} = \langle \phi_\mathrm{nematic}^4 \rangle/\langle \phi_\mathrm{nematic}^2 \rangle^2$, where
\[\phi_\mathrm{nematic} = \frac{1}{\sqrt{\beta L^2}} \int_0^\beta d\tau \left(\sum_{\langle ij\rangle_y} \tau^z_{ij}(\tau) - \sum_{\langle ij\rangle_x} \tau^z_{ij}(\tau)\right).\]
Here we use isotropic hopping parameters, $V=0.5$ and consider different values of $h$. For $L \geq 10$, the curves cross at $h \approx 3.19$, and moreover, there is no peak growing with $L$ near the transition.

\begin{figure}[!htb] 
  \includegraphics[width=5.0in]{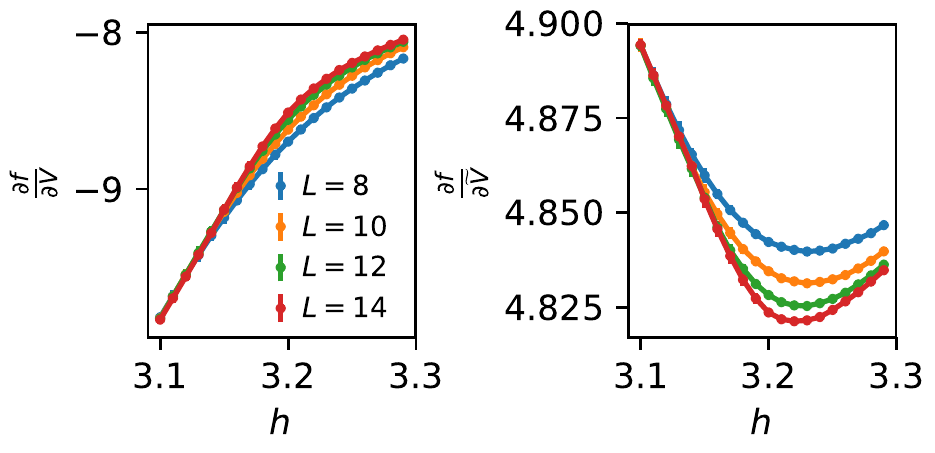}
  \caption{First derivatives of free energy density for various system sizes, across a transition from NAF to AF, with $\beta=10.0$. There does not appear to be a discontinuous jump at the transition $h_c \approx 3.19$.}
  \label{derivative-naf-to-af}
\end{figure}

We can also directly examine the first derivatives of the free energy density $f$. Two such derivatives which demonstrate the biggest changes near the transition are illustrated in Fig. \ref{derivative-naf-to-af}, and we observe no signatures of a discontinuous jump.

\section{Valence-bond solid to antiferromagnet transition}
\begin{figure}[!htb] 
  \includegraphics[width=5.0in]{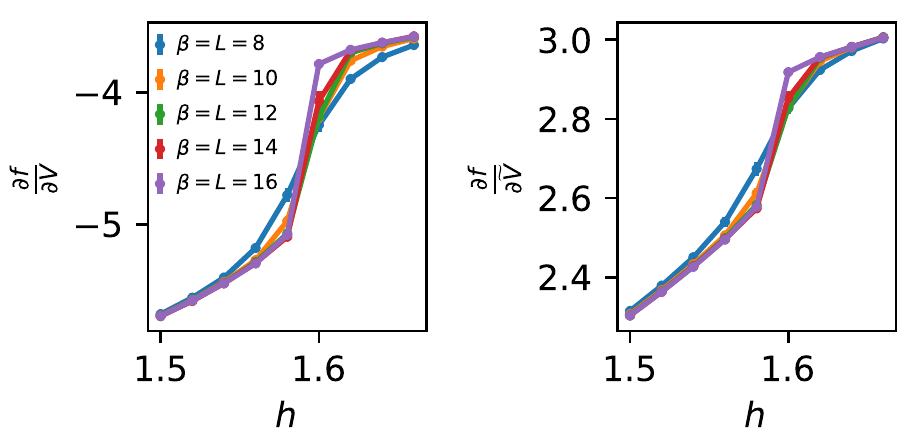}
  \caption{First derivatives of free energy density for various system sizes, across a transition from VBS to AF, with $\beta=L$. They exhibit a discontinuous jump at the transition near $h_c \approx 1.6$, which indicate that the transition is first order.}
  \label{jump-vbs-af}
\end{figure}

Here we provide information about the transition between the VBS and AF phases. We begin by considering a strongly first order transition, obtained by tuning $h$ at $V=0.25$ while maintaining $\beta=L$. Fig. \ref{jump-vbs-af} shows two first derivatives of the free energy density, which feature significant discontinuous jumps at the transition near $h_c \approx 1.6$, indicating that the transition is strongly discontinuous. The transition exhibits no larger emergent symmetries ($U(1)$ or $O(5)$).

We close by mentioning that by increasing the electron-phonon coupling $\alpha$ to $\alpha=1.5$, we have found a transition from VBS to NAF which is more weakly first order, but also without any emergent symmetries.

\end{bibunit}